\documentclass{jetpl}
\twocolumn
\input{psfig}

\lat

\newcommand\ket[1]{\left| #1\right\rangle}

\title{Dephasing of qubits by transverse low-frequency noise}

\rtitle{Dephasing by transverse low-frequency noise}

\sodtitle{Dephasing of qubits by transverse low-frequency noise}

\author{Yu.\,Makhlin$^{+,*}$\/\thanks{e-mail: 
makhlin,shnirman@tfp.uni-karlsruhe.de} and A.\,Shnirman$^+$}

\rauthor{Yu.\,Makhlin, A.\,Shnirman}

\sodauthor{Makhlin\,Yu., Shnirman\,A.}

\address{$^+$Institut f\"ur Theoretische Festk\"orperphysik,
Universit\"at Karlsruhe, D-76128 Karlsruhe, Germany\\
$^*$L.D.Landau Institute for Theoretical Physics RAS,
117940 Moscow, Russia}

\dates{September 23, 2003}{*}

\abstract{We analyze the dissipative dynamics of a two-level quantum system 
subject to low-frequency, e.g. $1/f$ noise, motivated by recent experiments with 
superconducting quantum circuits. We show that the effect of transverse linear 
coupling of the system to low-frequency noise is equivalent to that of quadratic 
longitudinal coupling. We further find the decay law of quantum coherent 
oscillations under the influence of both low- and high-frequency fluctuations, 
in particular, for the case of comparable rates of relaxation and pure 
dephasing.}


\begin{document}

\maketitle

Recent experiments with superconducting Josephson-junction 
circuits~\cite{Saclay_Manipulation_Science,Exps} demonstrated quantum coherent 
oscillations with a long decay time and a quality factor up to $\sim 10^4$. 
These experiments, on one hand, probe coherent properties of Josephson qubits 
(quantum bits) and demonstrate their potential for applications in quantum 
computing and quantum communication. On the other hand, they may be viewed as a 
probe of the noise mechanisms in the devices studied.

For the description of the dynamics of a two-level system (qubit, spin) subject 
to weak short-correlated noise one may use the Bloch equation, known from the 
NMR studies, which describes exponential relaxation of the longitudinal spin 
component and dephasing of the precessing transverse spin component (here and 
below we use the spin-1/2 language to discuss the dynamics). This description is 
valid as long as the correlation time of the noise is short compared to the 
typical dissipative times $T_1$, $T_2$. However, in Josephson-junction qubits 
the low-frequency noise is strong. These low-frequency fluctuations are 
correlated over distant times, and special treatment of their influence on a 
qubit is needed. They could lead to complicated decay 
laws~\cite{Cottet_Naples,Nakamura_Echo,Shnirman_Makhlin_Schoen_Nobel,%
Paladino_1/f}. In Ref.~\cite{Our_X2} the influence of low-frequency fluctuations 
nonlinearly coupled to a qubit was analyzed; this analysis is relevant for 
operation at the so-called optimal operation 
points~\cite{Saclay_Manipulation_Science}. Here we extend this analysis to 
account for the effect of transverse fluctuations also present at optimal 
points. While our discussion applies to an arbitrary dissipative two-level 
system, for illustration we consider the Josephson charge qubit, similar to that 
studied in the experiment~\cite{Saclay_Manipulation_Science}. We begin by 
discussing this system and the relevant noise sources and then proceed to the 
analysis of dephasing in general and at optimal points.

%%%%%%%%%%%%%%%%%%%%%%%%%%%%%%%%%%%%%%%%%%%%%%%%%%%%%%
\begin{figure}
\centerline{\hbox{\psfig{figure=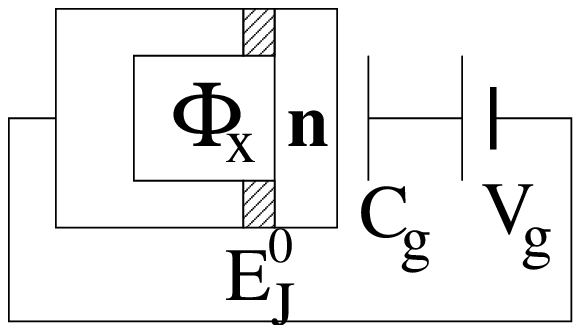,width=0.55\columnwidth}}}
\caption[]{\label{F:qb}FIG.\ref{F:qb}.
The simplest Josephson charge qubit}
\end{figure}
%%%%%%%%%%%%%%%%%%%%%%%%%%%%%%%%%%%%%%%%%%%%%%%%%%%%%%

\noindent{\bf Dissipative dynamics of a Josephson charge qubit}.  The simplest
Josephson charge qubit is the Cooper-pair box shown in
Fig.\ref{F:qb}~\cite{Our_RMP}.  It
consists of a superconducting island connected by a dc-SQUID (effectively, a 
Josephson junction with the coupling $E_{\rm J}(\Phi_{\rm x}) = 2E_{\rm 
J}^0\cos(\pi\Phi_{\rm x}/\Phi_0)$ tunable via the magnetic flux $\Phi_{\rm x}$; 
here $\Phi_0=hc/2e$) to a superconducting lead and biased by a gate voltage 
$V_{\rm g}$ via a gate capacitor $C_{\rm g}$. The Josephson energy of the 
junctions in the SQUID loop is $E_{\rm J}^0$, and their capacitance $C_{\rm 
J}^0$ sets the charging-energy scale $E_C \equiv e^2/2(C_{\rm g}+C_{\rm J})$, 
$C_{\rm J}= 2 C_{\rm J}^0$.  At low enough temperatures single-electron 
tunneling is suppressed and only even-parity states are involved. Here we 
consider low-capacitance junctions with high charging energy $E_C \gg E_{\rm 
J}^0$. Then the number $n$ of Cooper pairs on the island (relative to a neutral 
state) is a good quantum number; at certain values of the bias $V_{\rm g}\approx 
V_{\rm deg}= (2n+1)e/C$ two lowest charge states $n$ and $n+1$ are 
near-degenerate, and even a weak $E_{\rm J}$ mixes them  strongly. At low 
temperatures and operation frequencies higher charge states do not play a role. 
The Hamiltonian reduces to a two-state model,
\begin{equation}
         {\cal H} = - \frac{1}{2} [E_{\rm ch}(V_{\rm g}) \hat\sigma_z 
        + E_{\rm J}(\Phi_{\rm x})\hat\sigma_x] \,,
\label{MF}
\end{equation}
in the basis $\ket{\downarrow} = \ket{n}$, $\ket{\uparrow} = \ket{n+1}$; 
here $E_{\rm ch}(V_{\rm g}) =2e(V_{\rm g}-V_{\rm deg})
C_{\rm g}/(C_{\rm J} + C_{\rm g})$. The Hamiltonian (\ref{MF}) can be controlled 
via the gate voltage $V_{\rm g}$ and the applied flux $\Phi_{\rm x}$, this 
allows one to manipulate the qubit's state and perform quantum logic operations. 
To read out the final quantum state one has to couple the qubit to a quantum 
detector, e.g. a single-electron transistor~\cite{Our_RMP}.

Quantum bits are inevitably coupled to fluctuations in the environment (bath).
This destroys the coherence of the qubits' dynamics.  To slow down the dephasing
the coupling should be made as weak as possible.  In solid-state systems
decoherence is potentially strong due to numerous microscopic modes.  In 
Josephson qubits the noise is dominated by material-dependent
sources, such as background-charge fluctuations or variations of magnetic fields 
and critical currents, with power spectrum peaked at low frequencies,
often $1/f$.  A further relevant contribution is the electromagnetic noise of
the control circuit, typically Ohmic at low frequencies.  The $1/f$ noise
appears difficult to suppress and, since the dephasing is dominated by
low-frequency noise, it is particularly destructive.  On the other hand,
Vion~et~al.~\cite{Saclay_Manipulation_Science} showed that the effect of this
noise can be substantially reduced by tuning the linear longitudinal qubit-noise
coupling to zero (in a modified design; they also suppressed the coupling to the
quantum detector to minimize its effect on the qubit before the read-out).  This 
increased the coherence time by 2--3 orders of magnitude compared to earlier
experiments.

Of special interest is the analysis of the slow dephasing at such an optimal 
point. On one hand, comparison of theory and experiment may verify our 
understanding of the physics of the device studied as a dissipative two-level 
system. Further, from the analysis of the dephasing time scale and the decay law 
one may extract additional information about the statistical properties of the 
noise. On the other hand, understanding of the dissipative processes should 
allow their further suppression in future qubit designs.

Part of the noise (incl.~the background-charge fluctuations) can be thought of 
as fluctuations of the gate voltage and another part as fluctuations of the 
control flux $\Phi_{\rm x}$. It is convenient to discuss the effect of, e.g., 
the voltage noise $V_{\rm g}=V_{\rm g}^0+Y(t)$ in the qubit's eigenbasis:
\begin{equation}
{\cal H}= -\frac{1}{2} [\Delta E\; \hat \sigma_z + \zeta \hat Y(t) (-\sin\eta\; 
\hat \sigma_x + \cos\eta\; \hat \sigma_z)]\,,
\label{DisY}
\end{equation}
where the level splitting $\Delta E=(E_{\rm ch}(V_{\rm g}^0)^2 +
E_{\rm J}^2)^{1/2}$ and the angle between the static and fluctuating `magnetic' 
fields is given by $\tan\eta=E_{\rm J}/E_{\rm ch}(V_{\rm g}^0)$. We expanded the 
variation of $E_{\rm ch}$ in $Y$ to the linear order. Consider first the effect 
of weak short-correlated noise (with correlation time shorter than the 
dissipative times; this includes the finite-temperature Ohmic noise). In this 
case one can use the lowest-order perturbation theory and finds that the spin 
dynamics is described by the Bloch equations, known from NMR. The interlevel 
transitions are induced by the {\it transverse} fluctuations $\propto\sin\eta$ 
and give the relaxation time $1/T_1 = \zeta^2\sin^2\eta S_Y(\omega=\Delta E)/2$; 
the dephasing time is $1/T_2 = 1/(2T_1) + 1/T_2^*$, where the pure dephasing is 
induced by the {\it longitudinal} noise $\propto\cos\eta$ and gives $1/T_2^* = 
\zeta^2\cos^2\eta S_Y(\omega=0)/2$ (here the noise power $S_{Y}(t)=(1/2)\langle 
[Y(t),Y(0)]_+ \rangle$; we set $\hbar=1$). The effect of the magnetic-flux noise 
can be analyzed similarly. For Josephson qubits these expressions give good 
estimates for the measured relaxation times but do not suffice to describe the 
dephasing. Indeed, the expression for $T_2^*$ cannot be used for strong 
longitudinal low-frequency, e.g. $1/f$ noise; still it indicates that dephasing 
is strong. In first experiments~\cite{Nakamura_Nature,Nakamura_Echo} dephasing 
times in the range of fractions to a few nanoseconds were achieved. Tuning to an 
optimal point extended the coherence time to 
$\sim1\,\mu$s~\cite{Saclay_Manipulation_Science}.

\noindent{\bf Dephasing at optimal point}.
We illustrate our discussion of decoherence at an optimal point by considering a 
qubit deep in the charge limit, although in the device of 
Ref.~\cite{Saclay_Manipulation_Science} $E_{\rm C}$ and $E_{\rm J}$ were 
comparable (in which case two lowest eigenstates, which form the qubit, are no 
longer charge states). Using two control parameters $V_{\rm g}$ and $\Phi_{\rm 
x}$ one can tune the longitudinal linear couplings to the charge and flux noise 
to zero: For instance, for the system (\ref{DisY}) tuning the gate voltage to 
the degeneracy point $E_{\rm ch}(V_{\rm g}^0)=0$ yields $\cos\eta=0$. Further, 
tuning $\Phi_{\rm x}$ to the point of maximal $E_{\rm J}(\Phi_{\rm x})$ also 
suppresses the linear coupling to the flux fluctuations $\Phi_{\rm x} = 
\Phi_{\rm x}^0 + X(t)$. Thus, at this optimal point the Hamiltonian reads:
\begin{equation}
{\cal H} = -\frac{1}{2}[\Delta E\hat\sigma_z + \lambda X^2 \hat\sigma_z + \zeta 
Y \hat\sigma_x
]\,,
\label{Eq:HamXY}
\end{equation}
where we left only the leading fluctuating terms.

The quadratic londitudinal low-frequency noise $\lambda X^2 \hat\sigma_z$ may 
result in an unusual dephasing law (with a power law crossing over to 
exponential decay) due to strong higher-order contributions~\cite{Our_X2}. Here 
we discuss the effect of the transverse noise $\zeta Y \hat\sigma_x$. It can 
lead to relaxation processes and contribute to pure dephasing in higher orders. 
Thus in the analysis of dephasing one needs to account for both $\lambda X^2$ 
and $\zeta Y$ terms.

The effect of the low-frequency ($\omega\ll\Delta E$) transverse noise can be 
treated in the adiabatic approximation: we diagonalize (\ref{Eq:HamXY}) to 
$-\hat\sigma_z\sqrt{(\Delta E+\lambda X^2)^2 + (\zeta Y)^2}/2  \approx
-\hat\sigma_z [\Delta E +\lambda X^2 + \zeta^2Y^2/(2\Delta E)]/2 $, thus the 
low-$\omega$ transverse noise contributes to pure dephasing. In general 
higher-frequency fluctuations are also present and induce relaxation. If the 
relaxation is much slower than the pure dephasing, one may neglect its 
contribution to the total dephasing. If the relaxation is much faster, it 
dominates the decoherence; in this limit its rate $\zeta^2 S_Y(\Delta E)/2$ is 
given by the golden rule. However, of special experimental 
interest~\cite{Saclay_Manipulation_Science} is the situation with comparable 
relaxation and pure-dephasing time scales. We analyze whether evaluation of each 
of them is influenced by the other in this case, that is whether the low- and 
high-$\omega$ contributions interfere. In particular, we expect~\cite{Our_X2} 
strong higher-order contributions to the pure dephasing due to strong 
low-$\omega$ noise. Does it also contribute to relaxation? In the lowest order 
the relaxation is due to transitions with emission of a single resonant bath 
excitation; can instead a near-resonant excitation be emitted accompanied by 
low-frequency excitations? Here we show how the dephasing and relaxation laws 
and time scales can be obtained.

\noindent{\bf Dephasing by transverse noise}.
We begin with a discussion of purely transverse noise, $\lambda=0$. We focus on 
the long-correlated noise (slow decay of $\langle Y(0)Y(t) \rangle$), i.e. on 
the noise power peaked at low and smooth at high frequencies. In our analysis 
below we assume weak dissipation: pure dephasing and relaxation slower than the 
osillations, $\Gamma\ll\Delta E$, where $\Gamma$ represents the total-dephasing 
time scale. This limit is of primary interest for the circuits that realize 
qubits. Further, below for illustration we consider a source of gaussian noise 
$Y(t)$, which can be characterized by its second correlator, but our major 
conclusions persist in more general situations.

Our discussion is based on the analysis of the evolution operator of the qubit 
dynamics using the `real-time' Keldysh diagrammatic expansion in the qubit-bath 
coupling (this approach~\cite{Schoeller_PRB} is useful, since the spin degree of 
freedom does not satisfy the Wick theorem; it reminds the approach of 
Ref.~\cite{Konstantinov_Perel}). We begin by showing that the subleading-order 
effects of the low-frequency transverse noise reduce to the lowest-order 
contribution of longitudinal quadratic noise (this can also be seen from the 
adiabatic approximation but our derivation indicates the diagrams, important in 
the discussion below).

\begin{figure}
\centerline{\hbox{\psfig{figure=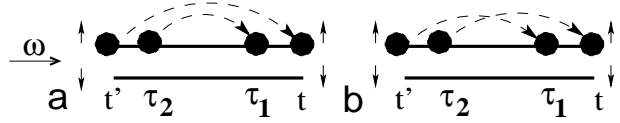,%
width=0.95\columnwidth}}}
\caption[]{FIG.~\ref{F:2order}. Second-order contributions to the self-energy 
$\Sigma_{\uparrow\downarrow \leftarrow \uparrow\downarrow}$. Other terms are 
obtained by shifting an even number of the vertices in {\bf a} or {\bf b} to the 
lower branch.}
\label{F:2order}
\end{figure}

In the diagrams the horizontal direction explicitly represents the time axis. 
The solid lines describe the unperturbed (here, coherent) evolution of the 
qubit's $2\times 2$ density matrix $\hat\rho$, $\exp(-iL_0t)\theta(t)$, where 
$L_0$ is the bare Liouville operator (this translates to $1/(\omega-i L_0)$ in 
the frequency domain). The vertices are explicitly time-ordered; each of them 
contributes the term $\zeta Y \sigma_x\tau_z/2$, with the bath operator $Y(t)$ 
and the Keldysh matrix $\tau_z=\pm1$ for vertices on the upper/lower time 
branch. Averaging over the fluctuations should be performed; for gaussian 
correlations it pairs the vertices as indicated by dashed lines in 
Fig.\ref{F:2order}, each of the lines corresponding to a correlator $\langle 
YY\rangle$.
Fig.\ref{F:2order} shows contributions to the 
second-order self-energy $\Sigma_{\uparrow\downarrow \leftarrow 
\uparrow\downarrow}^{(2)}$ (here $ij=\uparrow\downarrow$ label four entries of 
the qubit's density matrix). The term in Fig.\ref{F:2order}a gives
\begin{equation}
\label{Eq:2order_time}
\left(\frac{\zeta}{2}\right)^4\!\!\!\!
\int\!\! d\tau_1 d\tau_2 \langle Y(t)Y(t')\rangle
\langle Y(\tau_1)Y(\tau_2)\rangle
e^{i\Delta E(\tau_1 - \tau_2)}
\,,
\end{equation}
with integration over the domain $t'<\tau_2<\tau_1<t$.

%%%%%%%%%%%%%%%%%%%%%%%%%%%%%%%%%%%%%%%%%%%%%%%%%%%%%%%%%%%%%
%%%%%%%%%%%%%%%%%%%%%%%%%%%%%%%%%%%%%%%%%%%%%%%%%%%%%%%%%%%%%

After the summation over vertex positions on the lower/upper branches in 
Figs.\ref{F:2order}a,b, we evaluate the behavior of the Fourier-transformed 
self-energy in the
vicinity of the level splitting, at $\omega=-\Delta E -\omega' +i0$, where 
$\omega' \ll \Delta E$. If the integral is dominated by low frequencies, 
we find:
\begin{eqnarray}
&&{\rm Re}\,\Sigma_{\uparrow\downarrow \leftarrow \uparrow\downarrow}^{(2)}
(\omega=-\Delta E - \omega' +i0 )
 \approx -
\frac{\zeta^4}{8\Delta E^2}
\int\frac{d\nu}{2\pi}
\nonumber\\
&&\times
\left[\langle Y^2_{\nu+\omega'}\rangle \langle Y^2_{-\nu}\rangle + 
\langle Y^2_{\nu-\omega'}\rangle \langle Y^2_{-\nu}\rangle\right]
=-\frac{1}{2} S_{Y_2}(\omega')\,,
\label{Eq:2order_BR}
\end{eqnarray}
where $S_{Y_2}$ is the noise of
\begin{equation}
Y_2 \equiv \zeta^2 Y^2/(2\Delta E)\,.
\end{equation}
The result (\ref{Eq:2order_BR}) coincides, as expected, with the lowest-order 
contribution of the term $-Y_2\;\sigma_z/2$; note that the left and right vertex 
pairs in Fig.\ref{F:2order} can be viewed as composite vertices corresponding to 
$-Y_2\;\sigma_z/2$. Below we demonstrate that similar reduction occurs in every 
order of the perturbative expansion. Specifically, we show that the decay of the 
off-diagonal entry of the density matrix is $\rho_{\uparrow\downarrow}(t) = 
\rho_{\uparrow\downarrow}(0) \cdot \exp(-t/2T_1) \cdot \gamma_{\rm \varphi}(t)$, 
where the relaxation time is given by the golden rule and the pure dephasing 
term $\gamma_{\rm \varphi}(t)$ is the same as for the longitudinal fluctuations 
$-Y_2\;\sigma_z/2$ (analyzed in~Ref.~\cite{Our_X2}).

To demonstrate this we consider the diagrammatic calculation of the evolution 
operator for the density matrix. We begin by evaluating the evolution of the 
off-diagonal entry $\rho^\uparrow_\downarrow$ (the phase dynamics and 
dephasing), and then discuss relaxation (evolution of $\rho^\uparrow_\uparrow$, 
$\rho^\downarrow_\downarrow$).

The expansion of the propagator contains vertices on the horizontal solid lines, 
representing the Keldysh contour. For a given number and ordering of the 
vertices, one is to integrate over their time positions, and then add up all 
diagrams. Consider the dependence of the integrand on the time position of a 
vertex. This dependence includes fast oscillations with frequency $\pm\Delta E$, 
since the vertex flips the spin and changes the energy of the bare hamiltonian 
$-\Delta E\;\sigma_z/2$, and a much slower dependence of the dashed line. Thus 
the integrand is a fast oscillating function of the time positions of each 
vertex. Since integration is typically performed over time range much wider than 
the period of oscillations (at all times of interest for the analysis of 
dephasing; this range is $\sim 1/\Gamma$ at $t \sim 1/\Gamma$), the contribution 
of the most part of the integration space (with vertices' time positions as 
coordinates) is strongly suppressed by fast oscillations. However, in certain 
directions in this space, in which pairs of vertices with opposite oscillation 
frequencies $\pm\Delta E$ move together, the variation is slow, and the 
respective domains dominate the integral. One can arrive at this conclusion, and 
determine the dominant domains, by considering the evaluation of a particular 
diagram: the integral over the time $t$ of a vertex $\int\nolimits_a^b g(t) 
\exp(i\Delta E t) dt$ is taken between the positions $a$, $b$ of the neighboring 
vertices. Since $g(t)$ is slow on scale $1/\Delta E$, the oscillatory integral 
is dominated by the boundary terms, $g(t)\exp(i\Delta E t)|_a^b/(i\Delta E)$. 
One can say that the vertex $t$ joins one of its neighbors, and later one 
integrates over the vertex-pair position, $a$ or $b$.

\begin{figure}
\centerline{\hbox{\psfig{figure=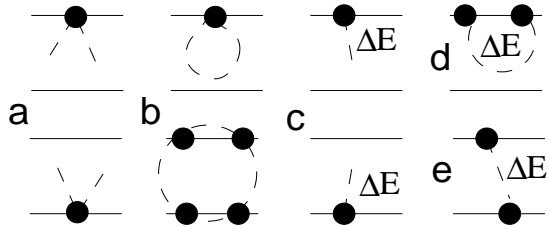,%
width=0.85\columnwidth}}}
\caption[]{FIG.~\ref{F:slow}. {\bf a.} Double vertices with low-$\omega$ tails, 
which appear in the evaluation of dephasing. {\bf b.} Examples of clusters built 
out of them~\cite{Our_X2}. {\bf c}. A low-$\omega$ object with a high-frequency 
dashed line. The relaxation process in {\bf e} also contributes to dephasing as 
shown in {\bf d}.}
\label{F:slow}
\end{figure}

One can continue this process, integrating at each step over time positions of 
unpaired vertices (or clusters with an odd number of vertices and hence 
oscillatory behavior), if any are still present. Finally one arrives at a 
situation where all vertices are paired, and the dependence of the integrand on 
pair time positions is slow (the exponentials $\exp(\pm i \Delta E t)$ for two 
paired vertices compensate each other). The integral in each domain with a fixed 
time ordering is dominated by the boundary terms, that is the terms with paired 
vertices. Thus we eliminate the high-frequency ($\sim\Delta E$) behavior, and 
now can evaluate the propagators using the diagram technique with ingredients 
that are slow (without oscillatory dependence on their time position): `double' 
vertices with two dashed tails in Fig.\ref{F:slow}a, and dashed lines connecting 
these tails (cf.~the examples in Fig.\ref{F:slow}b). Although 4-, 6- and further 
$2n$-fold clusters also form slow objects, their creation requires additional 
constraints on the vertex times (compared to building $n$ pairs) and the 
respective integration domain is much smaller; thus the contribution of such 
clusters is of higher order in $\Gamma/\Delta E$.

%%%%%%%%%%%%%%%%%%%%%%%%%%%%%%%%%%%%%%%%%%%%%%%%%%%%%%%%%%%%%
%%%%%%%%%%%%%%%%%%%%%%%%%%%%%%%%%%%%%%%%%%%%%%%%%%%%%%%%%%%%%

A closer inspection of the spin dependence and the Keldysh two-branch structure 
reveals that in each pair both vertices are located on the upper or both on the 
lower branch (for vertices located on different branches, two terms with 
different time orderings cancel each other unless the vertices are linked by a 
dashed line; such a term appears in the analysis of relaxation but not of the 
dephasing), and they indeed effectively correspond to the term 
$-Y_2\;\sigma_z/2$ in the Hamiltonian.

So far we constructed slow composite objects paying attention only to the 
oscillations of the solid lines in the diagrams and assuming very slow dashed 
lines, i.e., neglected the higher-frequency noise. In fact, one can construct 
another slow object shown in Fig.\ref{F:slow}c, if the respective oscillations 
of the solid lines are compensated by the dashed line from this vertex. In other 
words, in the frequency domain, one constrains the frequency of the dashed line 
to be $\Delta E$ (or $-\Delta E$, depepnding on the direction of the spin flip 
at the vertex). The dashed lines from such objects pair up, and the integral 
w.r.t. their relative position is dominated by small separations, $\delta t\sim 
1/\Delta E$. Thus one finds the slow object of Fig.\ref{F:slow}d, two vertices 
linked by a dashed line at frequency $\Delta E$; it describes the relaxational 
contribution to dephasing $\exp(-t/2T_1)$, where \begin{equation}
1/T_1=\zeta^2 S_Y(\Delta E)/2\,.
\label{Eq:Grel}
\end{equation}
In similar clusters of higher order additional constraints strongly limit the 
integration domain. Note that the object in Fig.\ref{F:slow}d involves the weak 
noise at a high frequency $\Delta E$, unlike those in Fig.\ref{F:slow}a, but it 
is still relevant since the lowest-order term in the upper part of 
Fig.\ref{F:slow}b is imaginary and does not contribute to dephasing. 

Similarly, we analyze the relaxation of the diagonal entries 
$\rho^\uparrow_\uparrow$, $\rho^\downarrow_\downarrow$. The new slow ingredient 
in this analysis is shown in Fig.\ref{F:slow}e. As for the composite objects in 
Fig.\ref{F:slow}a, the terms with these objects located on the upper and lower 
branches cancel each other, due to different signs ascribed to them in the 
Keldysh formalism (in contrast, in the analysis of the evolution of 
$\rho^\uparrow_\downarrow$ displacing a vertex from one branch to the other 
flips the spin thus yielding an additional sign change rather than 
cancellation). Hence the relaxation is given by the terms in Fig.\ref{F:slow}d,e 
and Eq.~(\ref{Eq:Grel}). We find that the strong low-frequency noise does not 
influence the relaxation rate.

{\bf Discussion}. We focused on the effect of the purely transverse noise. One 
can verify that in the presence of the longidsutinal fluctuations $-\lambda 
X^2\sigma_z/2$ the reduction persists: the relaxation is still given by 
Eq.~(\ref{Eq:Grel}), and the dephasing can be found by considering the 
longitudinal noise $-[\lambda X^2 + \zeta^2 Y^2/(2\Delta E)]\sigma_z/2$. For 
uncorrelated fluctuations $X(t)$ and $Y(t)$ their effects just add up (this 
would happen in the charge limit $E_{\rm C}\gg E_{\rm J}$ for the qubit in 
Fig.\ref{F:qb} at the degeneracy point, where the charge noise is transverse, 
and at the proper flux bias, where the flux noise is quadratic longitudinal). In 
the experiment~\cite{Saclay_Manipulation_Science} $E_{\rm C}$ and $E_{\rm J}$ 
were comparable, hence both charge and flux noise contributed to longitudinal 
and transverse fluctuations making them correlated; this should be taken into 
account but does not complicate the analysis.

We considered slow fluctuations. For short-correlated noise the double vertices 
of Fig.\ref{F:slow}a do not contribute, and one recovers the Bloch equations.

Furthermore, we illustrated our analysis by an example of gaussian noise. Such 
fluctuations are indeed encountered: The low-frequency noise, e.g. the 
background-charge fluctuations in Josephson circuits, is possibly produced by a 
collection of bistable fluctuators (or discrete system with more states). With a 
proper wide distribution of their parameters (couplings to the qubit, switching 
rates) they produce a smooth $1/f$ noise power. If the qubit is affected by many 
fluctuators, with a dense distribution in the parameter space, due to the 
central limit theorem one expects gaussian noise. In some Josephson devices 
sharp noise features indicate that a few fluctuators dominate and the resulting 
noise is non-gaussian (dephasing by such fluctuators was studied, e.g., in 
Ref.~\cite{Paladino_1/f}). We emphasize that our reduction applies also to the 
analysis of these systems. Indeed, our derivation used only the fact that the 
fluctuations are slow. Thus one can still build the diagrams from the slow 
objects constructed above but for non-gaussian fluctuations the dashed tails of 
the vertices may join not in pairs but also in larger bunches. After the 
reduction to quadratic longitudinal noise one may use other, non-diagrammatic, 
ways to analyze its effect.

Our results are relevant for the analysis of the experiment of 
Ref.~\cite{Saclay_Manipulation_Science}, in which the measured relaxation and 
dephasing times were comparable. The prediction of a specific decay law requires 
a detailed knowledge of the noise power of charge and flux fluctuations. It can 
be acquired via measurements away from the optimal point, as indicated in 
Ref.~\cite{Saclay_Manipulation_Science}.

We thank G.~Sch\"on and J.~Schriefl for useful discussions. This work is part of 
the CFN of the DFG. Y.M. was supported by the Humboldt foundation, the
BMBF, and the ZIP programme of the German government.

\end{document}